\documentclass[12pt,a4paper]{article}
\usepackage{graphicx}
\usepackage{times}
\title{\bf The Expanded Very Large Array}
\author{S.M.~Dougherty$^{1}$ and Rick~Perley$^2$\
\vspace{1cm}\\
\normalsize $^1$NRC-HIA, Penticton, Canada \\
\normalsize $^2$National Radio Astronomy Observatory, Socorro, NM, USA.}
\date{\mbox{}}
\begin{document}
\maketitle
\pagestyle{empty}
%
%
\def\bull{\vrule height .9ex width .8ex depth -.1ex}
\makeatletter
\def\ps@plain{\let\@mkboth\gobbletwo
\def\@oddhead{}\def\@oddfoot{\hfil\tiny\bull\quad
``The multi-wavelength view of hot, massive stars''; 39$^{\rm th}$ Li\`ege Int.\ Astroph.\ Coll., 12-16 July 2010 \quad\bull}%
\def\@evenhead{}\let\@evenfoot\@oddfoot}
\makeatother
%
%
\def\beginrefer{\section*{References}%
\begin{quotation}\mbox{}\par}
\def\refer#1\par{{\setlength{\parindent}{-\leftmargin}\indent#1\par}}
\def\endrefer{\end{quotation}}
%
%
{\noindent\small{\bf Abstract:} The Very Large Array is undergoing a major
upgrade that will attain an order of magnitude improvement in continuum
sensitivity across 1 to 50 GHz with instantaneous bandwidths up to 8 GHz in
both polarizations. The new WIDAR correlator provides a highly flexible
spectrometer with up to 16 GHz of bandwidth and a minimum of 16k channels
for each array baseline. The new capabilities revolutionize the
scientific discovery potential of the telescope. Early science programs are
now underway. We provide an update on the status of the project and a
description of early science programs.  }
%
%
\section{Background}
The Very Large Array has played a leading role in radio astrophysics
over 30 years since it was completed in 1980. Since then, the
capabilities of the VLA have changed very little. In order to continue
the unparalleled scientific achievements of the telescope, a major
expansion of its capabilities is currently nearing completion that
improve radically the capabilities of the VLA. The key goals are to
attain an order of magnitude improvement in continuum sensitivity (two
orders in survey speed), complete frequency coverage from 1 to 50 GHz
with vastly increased spectroscopic capability and correlator
flexibility. Such improved specifications will greatly enhance the
scientific discovery potential of the telescope, particularly in four
science areas:

\begin{itemize}
\item the magnetic universe - the structure and topology of magnetic fields
\item the obscured universe - enable unbiased imaging surveys of
dust-enshrouded objects that are obscured at higher frequencies
\item the transient universe - enable rapid response to transient events, and
\item the evolving universe - tracking the formation and evolution of
objects in the universe, from stars to galaxies to magnetic fields.
\end{itemize}

The key drivers within these broad science themes demand
noise-limited, full-field imaging in all Stokes parameters, and
point-source sensitivities of a few micro-Jy in an hour, leading to
imaging dynamic ranges greater than $10^6$, and a wide-range of
spectroscopic ability.  To attain such demanding specifications,
several primary hardware areas have been upgraded:
\begin{itemize}
\item new broad-band receiver systems that provide continuous coverage
between 1 to 50 GHz in eight different frequency bands, and superior
sensitivity compared to the VLA systems (see Figure 1).
\item new front-end electronics to digitize four 2-GHz-wide (R and L)
frequency pairs at each antenna for a total of up to 8-GHz
instantaneous bandwidth per polarization.
\item new wide-band fiber-optic data transmission system to carry
16 GHz of signal bandwidth from each antenna to the correlator. This 
will eliminate stability and calibration challenges imposed by the
analogue wave-guides of the VLA e.g. 3~MHz ripple and related spectral
baseline instabilities.
\item a new, highly-flexible, wide-band, full polarization correlator
with at least 16k channels per baseline, and adjustable frequency
resolution between 2.0 MHz and 0.12~Hz, using 64 independently tunable
sub-bands, leading to an enormous range of potential correlator
configurations, especially important for spectroscopy. Additionally,
WIDAR has many specialized modes - e.g. phased-array, pulsar gating,
pulsar binning etc, that increase greatly its scientific utility.
\end{itemize}

The EVLA project started in 2001 and is now nearing completion. The
upgrade of all front-end electronics in the 27 antennas was completed
in July 2009 and the WIDAR correlator began commissioning operations
and early science in March 2010. The installation of the new receiver
systems is on-going and will be completed in late 2012, and marks the
completion of the EVLA. This \$$90$~Million project is funded jointly
by the US National Science Foundation (NSF), the Canadian National
Research Council (NRC), and CONACyT, Mexico.

\section{EVLA Performance}

The upgrades in the EVLA system provide substantial improvement over the
performance characteristics of the VLA (Figure 1 and Table~1)

\begin{table}[!ht]
\caption[]{Comparison of overall EVLA and VLA performance characteristics}
\smallskip
{\small
\begin{center}
\begin{tabular}{|l|l|l|l|}
\hline
Parameter & VLA & EVLA & Factor \\
\hline
Continuum Sensitivity ($1\sigma$, 9 hr) & 10~$\mu$Jy & 1~$\mu$Jy & 10 \\
Maximum bandwidth in each polarization   & 0.1 GHz & 8 GHz & 80 \\
Number of channels at maximum bandwidth & 16      & 16,384 & 1024 \\
Maximum number of frequency channels    & 512 & 4,194,304 & 8192 \\
Coarsest frequency resolution           & 50 MHz & 2 MHz & 25 \\
Finest frequency resolution             & 381 Hz & 0.12 Hz & 3180 \\
Number of full-polarization sub-correlators & 2 & 64 & 32 \\
log (frequency coverage over 1-50 GHz)  & 22\% & 100\% & 5 \\
\hline
\end{tabular}
\end{center}
}
\end{table}

\begin{figure}[!ht]
\begin{center}
\includegraphics[bb=93 3 711 360,
width=0.75\textwidth,clip=]{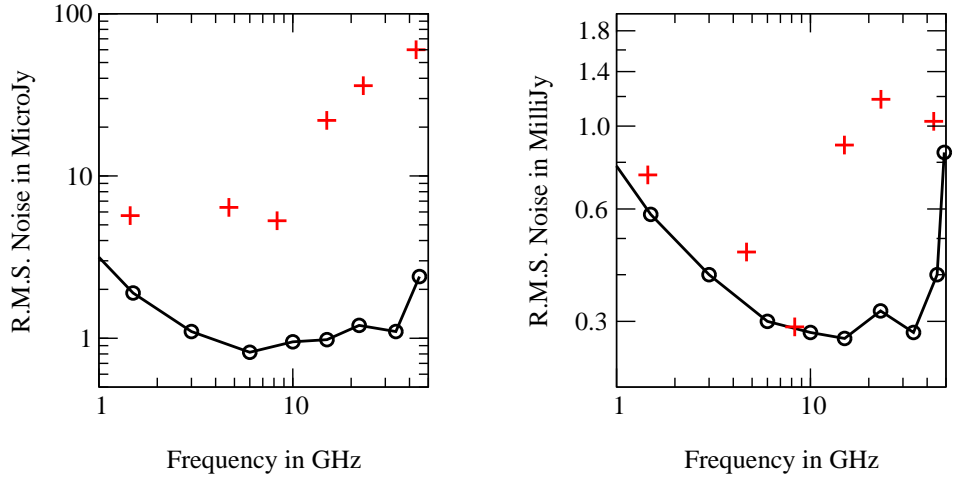}
\end{center}
\caption[]{The continuum (left) and line (right) sensitivity of the
EVLA (solid lines) compared to the VLA (crosses) ($1\sigma$ in 12
hours). For the continuum sensitivity, the full available bandwidth at
each band is assumed. For the spectral lines, the bandwidth adopted
corresponds to a velocity width of 1~km\,s$^{-1}$. Note the EVLA will
be continuously tunable over the entire frequency range.}
\end{figure}

A key element of the EVLA is the Wide-band Digital Architecture
(WIDAR) Correlator, designed and built by NRC-HIA in Penticton. The
major feature of this design is the correlation of the four 2-GHz IF
bands (giving 8-GHz of bandwidth in each polarization) via 64
independently tunable sub-band pairs with 16,384 channels per
baseline. There are 16 sub-band pairs associated with each 2-GHz IF
band. Each pair effectively forms an independent ``sub-correlator'',
and correlator configurations can be assigned to each pair independent
of the other pairs.  Each sub-band pair can have a sub-band width of
any of 128, 64, 32...,0.03125 MHz, and through recirculation the
number of spectral channels per baseline can be increased, using
certain correlator configurations, up to a maximum of 4.2M channels.

\subsection{Spectroscopy}
\label{subsec:spectroscopy}
The enormous flexibility in the configuration of the WIDAR correlator
resources enables the EVLA to meet or exceed the demands of spectroscopic
observations, and stands to revolutionize high-resolution centimetre
wave spectroscopy. 

An example of this flexibility is the simultaneous detection of multiple
spectral lines. It is possible to target up to 64 lines simultaneously, and
assign different spectral resolutions and sub-band widths for each sub-band
pair, if desired. Taking recombination line observing at S band (2-4 GHz)
as an example, there are 32 Hydrogen recombination lines that can all
targeted independently with 2~kHz resolution and 8 MHz sub-band width
(covering 1/8 the total bandwidth at S band), focusing correlator
resources on the spectral regions of interest. For weak lines, subsequent
stacking can be used to improve the signal-to-noise.  More extensive
examples of the efficiency of spectral line observing as a result of the
WIDAR sub-band design come at K band (18-26.5 GHz), of particular interest
to massive star research. Here, the EVLA could target the 32 molecular
density and temperature indicator lines with a velocity resolution of 0.2
km\,s$^{-1}$ e.g. including lines of NH$_3$, SO$_2$, H$_2$CS, H$_2$O,
H$_2$CO, CH$_3$OH, OCS (Figure 2). With the remaining resources, 8
sub-bands could be tuned to each of the Hydrogen recombination lines in
this frequency range, with the remaining 24 sub-bands covering
$24\times128$~MHz (3 GHz) of continuum. Another advantage of the wide
bandwidth, the continuum emission can be readily determined with the
abundance of ``empty'' channels.

There are innumerable variants on the configuration of the
WIDAR correlator, and potential users should consult the latest
operational status summary for the availability of correlator
configurations
(http://science.nrao.edu/evla/proposing/obsstatsum.shtml).

\begin{figure}[!ht]
\begin{center}
\includegraphics[angle=0,
width=0.75\textwidth,clip=]{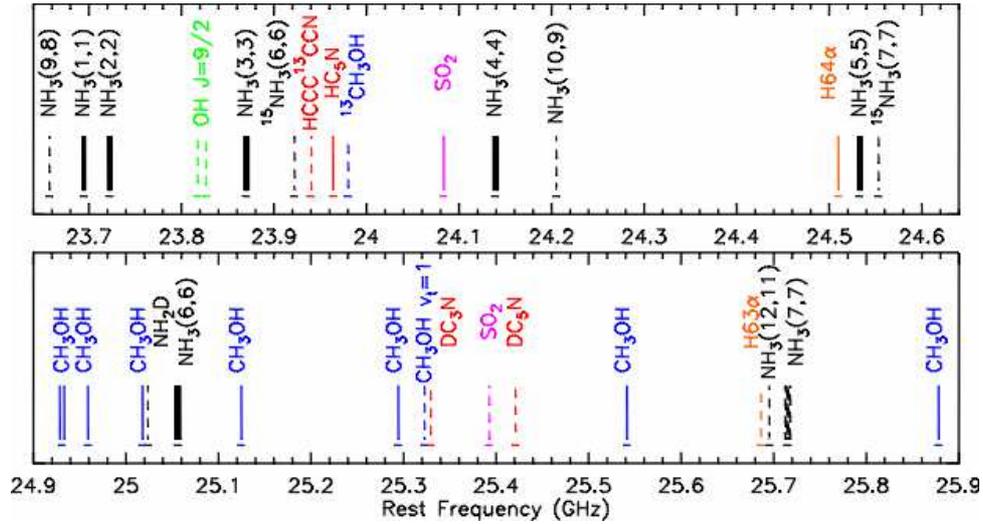}
\end{center}
\caption[]{A multi-line spectroscopic setup example that can be
attained with the WIDAR correlator, emplacing a sub-band at each of a
series of 32 lines at K band (18-26.5 GHz). This example is
taken from an RSRO experiment to study the conditions in a massive star
forming region (Figure 3).\label{fig:multi-line}}
\end{figure}

\begin{figure}
\begin{center}
\includegraphics[angle=0, width=0.75\textwidth,clip=]{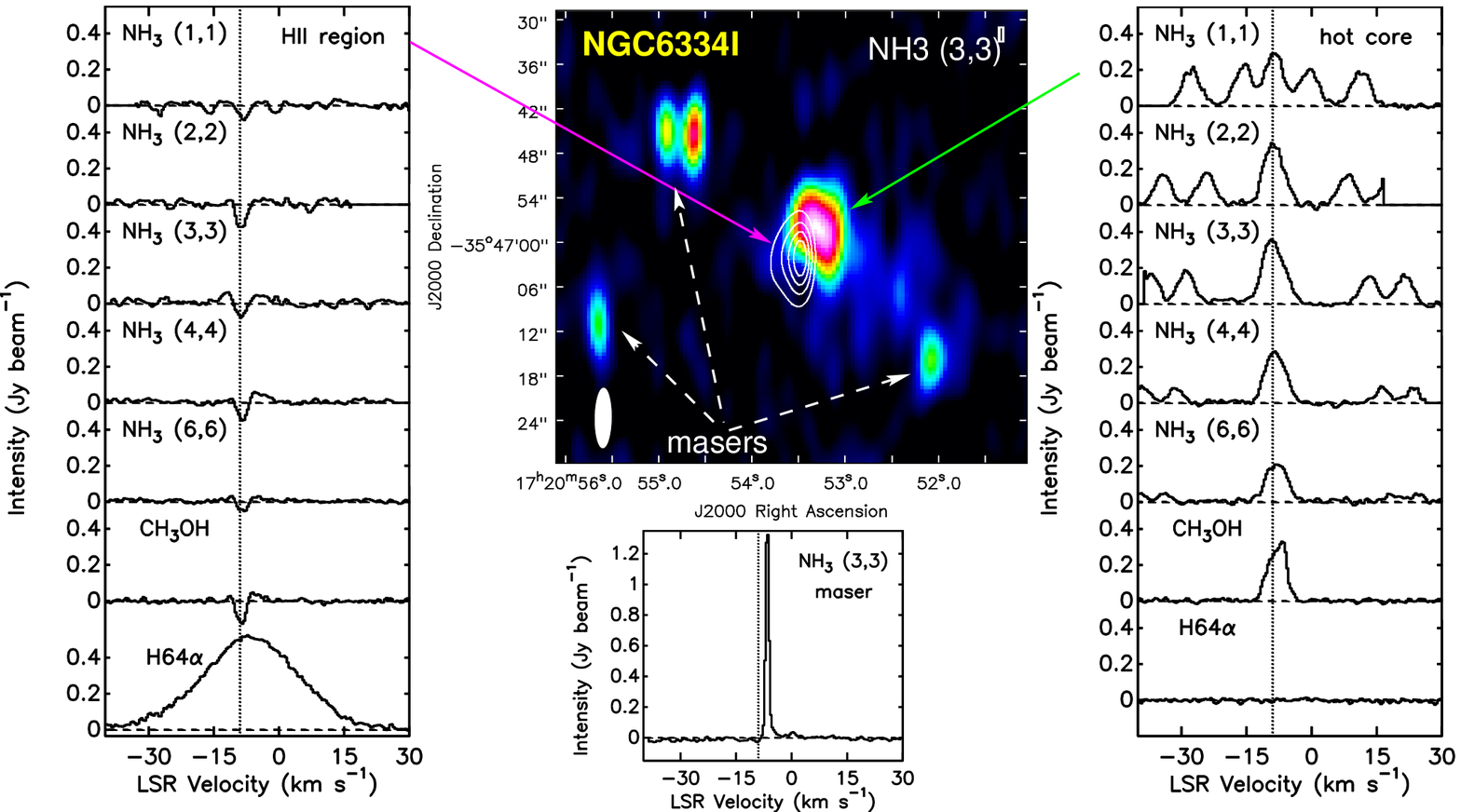}
\end{center}
\caption[]{An example of Early Science enabled through the RSRO
program, showing a K-band image of Ammonia in a massive star forming
region obtained in 10 minutes using $8\times8$~MHz sub-bands. The
contours are continuum emission from an H{\sc ii}
region. Ammonia in absorption against
the H{\sc ii} region (left panel) and in emission from one of the hot
cores (right panel). The panel at bottom-centre shows an ammonia maser
line.\label{fig:rsro_example}}
\end{figure}

\subsection{Continuum Imaging}
The digital data transmission system leads to improved calibration of
system gains, and together with the improved sensitivity of the new
receivers greatly enhances the continuum capability of the EVLA. Typical
image dynamic ranges used to be, at best, approaching $10^5$, requiring
major post-processing efforts. With the new system, dynamic ranges
approaching $10^6$ can be readily achieved with little effort, but reveal
that more sophisticated processing methods are required to handle the
residual image artifacts to attain the goal of greater than $10^6$.

The large bandwidth provides two immediate advances of relevance to
continuum emission studies. At full bandwidth the EVLA gives a factor of 10
improvement in sensitivity (0.1GHz vs 8 GHz), implying the potential to
detect continuum sources out to $3\times$ greater distance at the same
noise level, and hence flux limited sample sizes $\sim30\times$ larger than
available with the VLA. For massive stars, higher precision fluxes will lead
to more precise mass-loss estimates across the entire range of massive star
evolution. Another benefit is the vastly improved ability to determine the
continuum spectra as a result of the large number of potential
spectral fluxes that an be derived across each observing band and the
improved ability to match spectra across adjacent observing bands. For
stars, this will lead to improved understanding of, for example, the
nature of the underlying emitting particle energy spectrum in non-thermal
sources, and circumstellar envelope geometry of thermal emission sources.


\section{Early Science programs}
To make early and effective use of the new features of the EVLA as they are
commissioned, early science with the EVLA is enabled through two observing programs - the
Observatory-Shared Risk (OSRO) and Resident-Shared Risk (RSRO)
observations.

{\bf OSRO}: This program provides science capabilities that are similar to those
to the VLA, utilizing capabilities in the EVLA system that were
commissioned early in 2010, and is possible to access
``at-home''. Initially, up to 256 MHz of bandwidth is available
through two independently tunable 128-MHz bands of 64 channels each,
with both co- and cross-polarization products. For higher spectral
resolution, the total bandwidth of the sub-bands can be
reduced. Alternatively, 256 channels in one 128-MHz (or smaller)
sub-band is available. The available bandwidth in OSRO is expected to
grow in early 2011.

{\bf RSRO}: Science programs that require access to the more extensive
capabilities possible with the EVLA, particularly more sub-bands
and broader bandwidths, come under the auspices of the RSRO
program. These features are available to users in exchange for a
period of residence at the Array Operations Centre to aid in
commissioning.  The goal is to accelerate the development of the
scientific capabilities of the EVLA though the broad expertise of the user
community.  The current plan is for the RSRO program to run to the end
of 2011.  For those interested in participating in RSRO, a description
of the current status is available at
http://science.nrao.edu/evla/earlysicence/rsro.shtml.

\section{Summary}
Early science is already demonstrating the unprecedented potential of the
EVLA system, especially the flexibility and range of configurations of the
WIDAR correlator system. The science opportunities are tremendous and
ensure the EVLA will be the pre-eminent centimetre-wave radio telescope
over at least the next decade. Numerous challenges remain, particularly in
data processing and calibration, before the full diversity of science
goals can be achieved, but is it is through programs such as RSRO that
these challenges can be met in collaboration with the broad community.

%
%
%
\section*{Acknowledgments}
We would like to thank many people involved in the
design and commissioning of the EVLA in providing material for this
presentation. In particular, we thank Drs Crystal Brogan and Todd
Hunter for permission to present initial results from their RSRO science
program.

%
        
\end{document}